# Analytical quantitative description of wide plate U-free bend process of intelligent control

**SU Chun-jian WANG Wei-wei GUO Su-min**

(Collage of Mechanical and Electronic Engineering, Shandong University of Science and Technology, Qingdao, China 266590

e-mail: wangweiwei198701@gmail.com; suchunjian2008@163.com)

**Abstract:** According to flexure theory of plate, on the premise of plane deformation assumption, an analytic model of U-free bend theory is proposed in this paper, which considered the harden, anisotropy and elastic deformation of material. Then the theoretical analysis on U-free bending of wide plate is made, the change rules of bending force with bending stroke and formula of target angularity are obtained, the various factors affecting bending force and rebound are analyzed. Furthermore, theoretical calculation and experiment and simulation results are compared, which provided theoretical basis for identifying parameters of the wide plate U-free bending intelligent control as well as determining variables of the input layer and output layer of prediction model.

## 0 INTRODUCTION

The problem of rebound universally exists in the process of sheet forming, the phenomenon of rebound is more serious especially in the process of bending and light drawing , and it has a tremendous influence on dimensional accuracy of parts and production efficiency. So it is necessary to research rebound in depth and control it effectively. The final rebound shape of parts is a cumulative effect of its whole forming history, the process of sheet forming is closely associated with a number of factors, such as mold geometry, material characteristics, frictional contact etc, therefore the problem of rebound in the process of sheet forming is very complex. Many scholars at home and abroad researched the problem of rebound in depth, but all the studies are aimed at relatively simple V-free bending for analysis and discussion[1-3]. In this paper, an analytic model of comparatively complex U-shape flexure theory is established based on flexure theory of plate. Wide plate U-free bend theory is analyzed, the computing formula of bending moment, bending stroke, bending force and rebound angle are derived. And factors affecting sheet rebound and bending force are theoretically analyzed. Theoretical calculations and experiments and simulation results are compared, which provided theoretical basis for identifying parameters of the wide plate U-free bending intelligent control as well as determining variables of the input layer and output layer of prediction model[4-5].

## 1 Fundamental Assumption

In this paper, the following fundamental assumptions are adopted:

(1)The thinning on the direction of sheet thickness is ignored;

(2)The pure bending moment is supposed to exist in each section of the sheet length, so as to establish bending model by the flexure theory of pure bending moment segmentally;

(3)The external bending moment (Generated from the male punch and balanced by the internal bending moment) is assumed to be linear distribution from zero to the maximum, as shown in the Fig.1, the location of the maximum is point A, which is the contact position of the bottom of the punch and the sheet, the location of the minimum is point B, which is the contact position of the sheet and the female die, size zero;

(4)Three variable regions are included in the range of deformation: fully plastic bending in the contact area of the punch and the sheet; elastic-plastic bending in the range of $0 \le S \le S_E$ ; fully elastic bending in the range of $S_E \le S \le S_1$ .

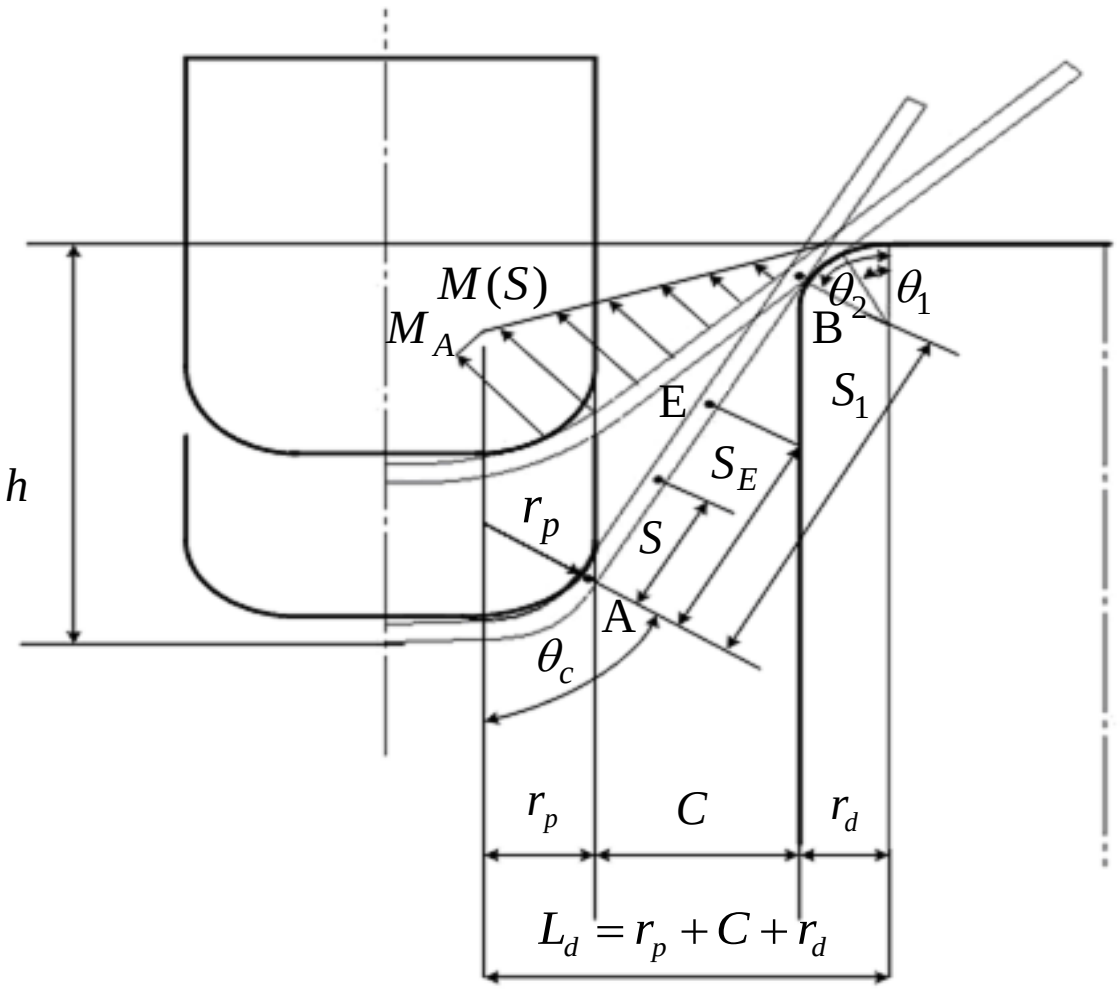

Fig.1 Geometrical relationship of U-free bending

## 2 Analysis of U-free bending

### 2.1 Analysis of bending process of U-free bending model

The procedure chart of U-bending deformation is shown in the Fig.2. The blank is bent between punch and die at the beginning of formation, as shown in the Fig.2(b). At this stage, the part of the figure between OA, is generally weighted with uniform bending moment and not contacted with the bottom surface of the punch, the part of the figure between AB is deformed along punch-nose radius. The bending force, in the process of bending, continued to increase with the increase of the punch stroke before the punch stroke reached $r_p + r_d$ , but the sheet would slip on the circular bead of female die and could pressed into the die easily when the stroke exceed $r_p + r_d$ , so the load dropped sharply, and the sheet continued to be pulled into the die under substantially unchangeable load. At this stage, the prominence generated at the bottom of the punch would almost has no change in the early stage of deformation, for the bending along with punch profile radius lied at the extremes of which has been completed for some time. When the bottom surface of the punch approach to the bottom surface of the die, the prominence generated at the bottom of the punch is flattened, so the bending force increases again, the result that the prominence been flattened made the sheet crashed out towards the direction of punch profile radius(Fig.2(c)). When the prominence is particularly remarkable, plates of the corner part of the punch left from the round part are changed into joggling shape. Generally, after the prominence being flattened, the deformation happened and makes the edge of the sheet folded towards the inside between OA and BC in Fig.2(c), and unfolded towards the outside between AB because of the elastic restitution of the sheet.

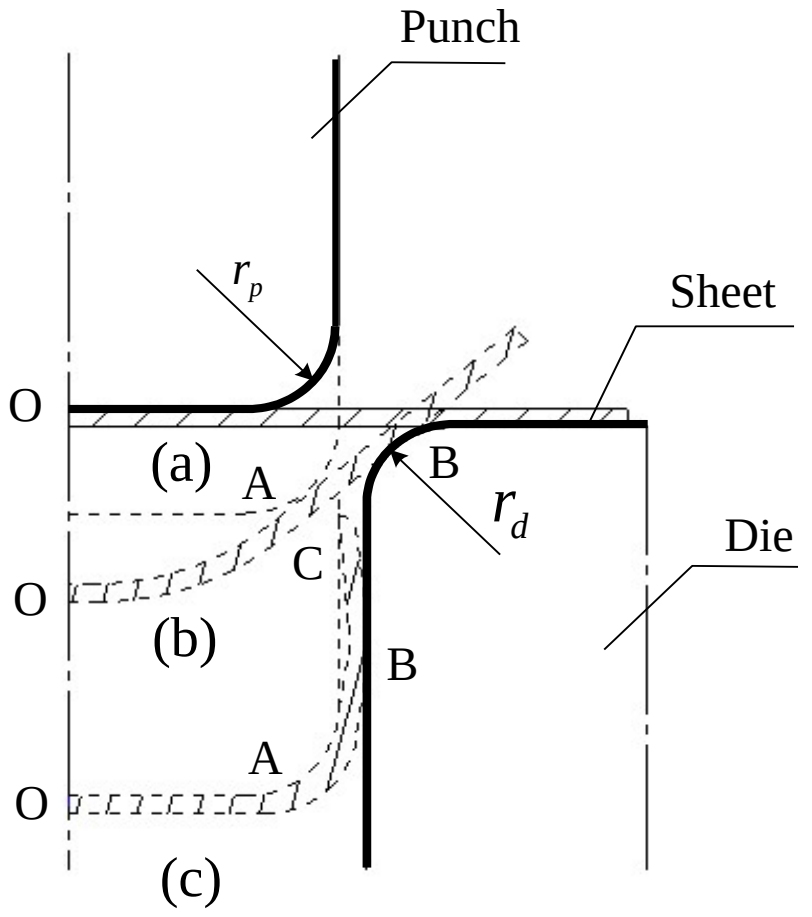

Fig.2 Stroke of U-bending

## 2.2 Calculation of bending moment

As shown in the Fig.1, point E can be seen as fully elastic bending, bending moment of which can be known:

$$M_E = \frac{wt^2}{6}\frac{\sigma_s}{1-\nu^2} \quad (1)$$

The sheet changes from elastic bending to elastic-plastic bending at the position of $S=S_E$, and a part of plastic yielding is caused by a certain number of internal moments balanced with the external moments. The bending moment and the plastic yielding augment gradually with the sheet to the direction of the punch, according to assumptions, the fully plastic bending would be reached at the contact area of punch and sheet, the internal bending radius of the sheet would be minimum(approximately equal to the punch bending radius), that is $r_i = r_p$. The elastic strain equals to zero in the point of $S=S_A=0$, or say the height of the elastic region is $2y_e = 0$ [6]. Then we have:

$$M_A = 2wkR^{n+1} r_n^2 e^{-2\varepsilon_0/R} \sum_{j=0}^{\infty} \{\frac{2^j - e^{\varepsilon_0/R}}{j!(j+1+n)}[(\varepsilon_{max} + \frac{\varepsilon_0}{R})^{j+1+n} - (\frac{\varepsilon_0}{R})^{j+1+n}]\}$$

In which:

$$\varepsilon_{max} = \ln\frac{r_0}{r_n} = \ln\frac{r_i + t}{\sqrt{r_i(r_i+t)}} = \ln\frac{\sqrt{1 + r_i/t}}{r_i/t} = \ln\frac{\sqrt{1 + r_p/t}}{r_p/t}$$

The elastic-plastic bending moment of point A can be obtained:

$$M_{ep} = C_4 r^2 + \frac{C_2}{r^n} + \frac{C_1}{r^{n+1}} \quad (2)$$

In which:

$$C_1 = \frac{3}{4}\frac{n+2}{n+3}C_2 t \text{ ; } C_2 = \frac{2wkR^{n+1}}{n+2}(\frac{t}{2})^{n+2} \text{ ;}$$

$$C_4 = \frac{2}{3}(\frac{\sigma_0}{E'})^2\sigma_0 w = \frac{2}{3}\varepsilon_{e,0}{}^2\sigma_0 w$$

The distribution of linear moment can be described as:

$$M(S) = M_E\frac{S_1 - S}{S_1 - S_E} \quad (3a)$$

Or:

$$M(S) = M_A\frac{S_1 - S}{S_1} = M_A(1 - \frac{S}{S_1}) \quad (3b)$$

The geometric relationship is as shown in the Fig.1, S is the distance from an arbitrary point on the sheet to point A. So the bending moment of the sheet's three models can be obtained:

$M(S) = M_A$ the contact area of the punch and the sheet(Fully plastic bending area) (4a)

$M(S) = M_A\frac{S_1 - S}{S_1} = M_{ep}$ $0 \le S \le S_E$ (Elastic-plastic bending area) (4b)

$M(S) = M_E\frac{S_1 - S}{S_1 - S_E}$ $S_E \le S \le S_A$ (Fully elastic bending area) (4c)

The sheet's critical distance $S_E$ changes from fully elastic bending to elastic-plastic bending can be obtained through the continuity of the bending moment in point A(S=0):

$$S_E = S_1(1 - \frac{M_E}{M_A}) = S_1 - \frac{1}{C_3} \quad (5)$$

In which: $C_3 = \frac{M_A}{M_E}\frac{1}{S_1}$

## 2.3 Calculation of curvature

The curvature is the function of the bending moment in the direction of the sheet length. There has a one-to-one corresponding relationship between curvature and bending moment. The curvature of each part of the sheet can be individually calculated through the equation (4(a))、(4(b))and(4(c))of the bending moment in the process of the bending formation. The sheet is wrapped around the punch in the contact area between punch and sheet, so the curvature can be seen as a fixed value in the contact area (fully plasticity) between punch and sheet:

$$\frac{1}{r} = \frac{1}{r_p + \frac{t}{2}} = \frac{1}{r_p'} \quad (6a)$$

In which: $r_p' = r_p + \frac{t}{2}$ is the radius of the sheet's central axis.

For the part of the elastic-plastic bending, the bending moments include elastic moment and plastic moment, the total moment is the equation (2), combined the equations (2) with (4), and according to the continuity

( $S=0$ , $M(0)=M_A$ ) of the bending moment at point A, we can obtain the following equation:

$$S_1=\frac{M_A}{C_3M_E}$$

Expression of the curvature distribution can be obtained:

$$S=S_1-C_6R^2-C_7/R^n-C_8/R^{n+1} \quad 0\le S\le S_E$$ (Elastic-plastic region) (6b)

In which:

$$C_6=\frac{S_1}{M_A}C_4\ ;\quad C_7=C_2\frac{S_1}{M_A}\ ;\quad C_8=\frac{3t}{4}\frac{n+2}{n+3}C_7$$

The curvature of the sheet's elastic bending area can be defined as following:

$$\frac{1}{r}=\frac{M(S)}{E'I}=\frac{M_E}{E'I}\frac{S_1-S}{S_1-S_E}=\frac{1}{r_E}\frac{S_1-S}{S_1-S_E}=\frac{d_3}{r_E}(S_1-S)$$

$$=C_5(S_1-S) \quad S_E\le S\le S_1$$ (Elastic region) (6c)

In which:

$$C_5=\frac{C_3}{r_E}=\frac{C_3M_E}{E'I}\ ;\quad E'=\frac{E}{1-v^2}\ ;\quad I=\frac{Wt^3}{12}$$

## 2.4 Calculation of angularity and rebound angle

The angularity (the total rotating angle) is the sum of the rotating angles of each part. The increment ( $dS$ ) of rotating radian can be defined as:

$$\mathrm{d}\theta=\frac{\mathrm{d}S}{r}$$

The increment of each part of rebound angle is:

$$\delta\theta_S=K_S\delta S=\frac{M(S)}{E'I}\delta S$$

The total rotating angle $\theta_{\text{zong}}$ can be obtained:

$$\theta_{\text{zong}}=\int_0^{\theta_2}\mathrm{d}\theta=\theta_C+\theta_{AE}+\theta_{EB} \tag{7}$$

In which: $\theta_C$ is the contact angle of punch and sheet, $\theta_{AE}$ is the rotating angle of the sheet bending elastic-plastic bending parts( $0\le S\le S_E$ ), $\theta_{EB}$ is the rotating angle of the sheet bending elastic bending parts( $S_E\le S\le S_1$ ).

According to the equation (7) and (6b, 6c), the following can be obtained:

$$\mathrm{d}\theta_{AE}=\frac{\mathrm{d}S}{r}=[-2C_6+nC_7(\frac{1}{r})^{2+n}+(1+n)C_8(\frac{1}{r})^{3+n}]\mathrm{d}r$$

$$\mathrm{d}\theta_{EB}=\frac{\mathrm{d}S}{r}=C_5(S_1-S)\mathrm{d}S$$

So the rotating bending angle of elastic-plastic parts and elastic parts can be obtained:

$$\theta_{AE}=\int_0^{S_E}\frac{\mathrm{d}S}{r}=\int_{r_p'}^{r_E'}[-2C_6+nC_7(\frac{1}{r})^{n+1}+(n+1)C_8(\frac{1}{r})^{n+2}]=$$

$$-2C_6(r_E'-r_p')+C_7[(\frac{1}{r_p'})^n-(\frac{1}{r_E'})^n]+C_8[(\frac{1}{r_p'})^{n+1}-(\frac{1}{r_E'})^{n+1}] \tag{8}$$

$$\theta_{EB}=\int_{S_E}^{S_1}\frac{\mathrm{d}S}{r}=\int_{S_E}^{S_p}C_5(S_1-S)dS=\frac{C_5}{2}(S_1-S_E)^2 \tag{9}$$

In which: $r'$ is the radius of neutral layer.

So the total rotating angle (angularity) can be obtained:

$$\theta_{\text{zong}}=\theta_C-2C_6(r_E'-r_p')+C_7[(\frac{1}{r_p'})^n-(\frac{1}{r_E'})^n]+$$

$$C_8[(\frac{1}{r_p'})^{n+1}-(\frac{1}{r_E'})^{n+1}]+\frac{C_5}{2}(S_1-S_E)^2 \tag{10}$$

Similarly, the total rebound angle $\theta_S$ is the sum of the rebound angles of each part:

$$\theta_S=\int_0^{S_1}\delta\theta_S=\int_0^{S_1}\frac{M(S)}{E'I}\delta S=\int_0^{S_1}\frac{C_3M_E}{E'I}(S_1-S)\delta S=$$

$$\frac{C_3M_E}{2E'I}S_1^2=\frac{M_A}{2E'I}S_1 \tag{11}$$

From the equation(11), one can see that: the rebound angle is mainly affected by bending length $S_1$ 、 bending moment and material property, the material property, mold shape and size, which affect the size of rebound angle, can be expressed by the equation(12):

$$\theta_S=\frac{1}{2}\frac{M_A}{E'I}S_1=3\frac{KR^{n+1}}{2+n}\frac{1-v^2}{E}(\frac{t}{2r_p'})^n(1+\frac{3}{2}\frac{n+2}{n+3}\frac{t}{2r_p'})\frac{S_1}{t} \tag{12}$$

The length $S_1$ of bending scope can be known from the Fig.1:

When $h\ge r_p+r_d+t$ , $S_1=\dfrac{(r_p+r_d+c)}{\cos\theta}-(r_p+r_d+t)\tan\theta$

## 2.5 Relationship between bending force and bending stroke

Fig.3 is a force diagram of wide plate U-free bending. Longitudinal balance equation is:

$$2\mu N\sin\theta+2N\cos\theta=P$$

After arrangement:

$$N=\frac{P}{2}\frac{1}{\cos\theta+\mu\sin\theta} \tag{13}$$

$$F=\mu N=\frac{P}{2}\frac{\mu}{\cos\theta+\mu\sin\theta} \tag{14}$$

In which: $P$ is bending force, $\mu$ is friction coefficient, $\theta$ is angularity.

The bending moment of point A can be expressed as:

$$M_A=F\frac{t}{2}+N\frac{h'}{\sin\theta}$$

According to geometrical relationship of Fig.3 and the equations (13) and (14), the external moment can be obtained:

$$M_A=\frac{p}{2(\mu\sin\theta+\cos\theta)}(\frac{\mu t}{2}+\frac{h'}{\sin\theta})$$

The expression of bending force can be obtained by making $M=M_{外}$ :

$$p=\frac{2M(\mu\sin\theta+\cos\theta)}{(\mu t/2+h'/\sin\theta)} \tag{15}$$

In which: $M$ is the bending moment generated by internal force.

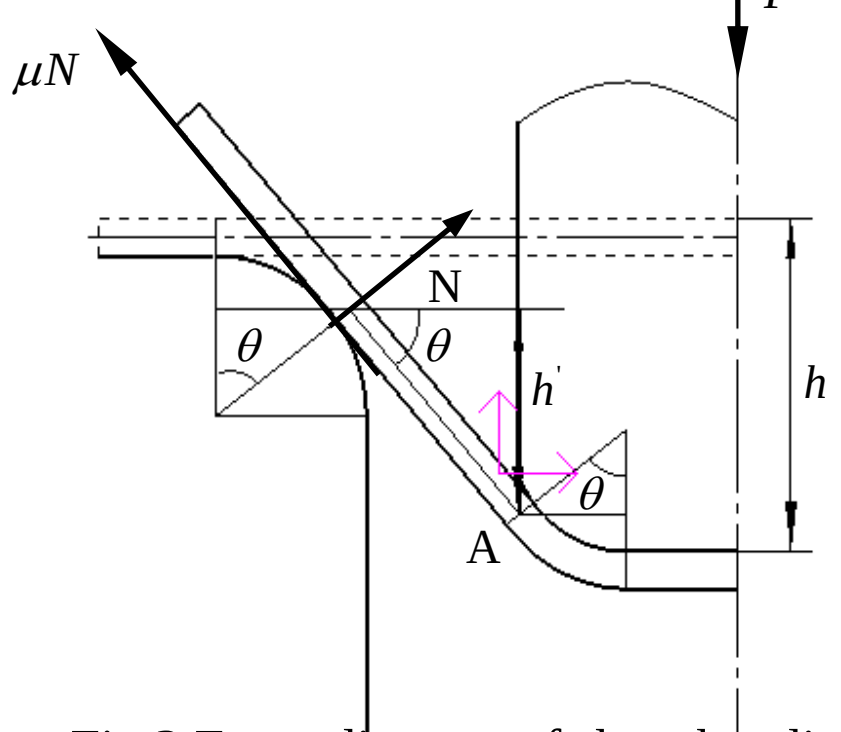

Fig.3 Force diagram of sheet bending

(1)While $h < r_p + r_d + t$, the relationship between the bending stroke $h$ and the angularity $\theta$, as shown in the Fig 4, can be obtained by geometrical relationship:

$$h' = h - \left(r_p + r_d + t\right)(1 - \cos\theta)$$

According to the geometrical relationship $CF = CD + DE + EF$ in the figure, we have:

$$[c + r_d + r_p(1 - \sin\theta)]\tan\theta = h - t - (r_p + r_d)(1 - \cos\theta) + r_d \sin\theta \tan\theta + t / \cos\theta$$

After arrangement:

$$h = [c + (r_p + r_d)(1 - \sin\theta)]\tan\theta + (r_d + r_p - t/\cos\theta)(1 - \cos\theta) \qquad (16)$$

When $h < r_p + r_d + t$, the sheet bends to a lesser degree, which can be seen as the elastic bending, the bending moment (M) is as in the equation (1), then:

$$P = \frac{wt^2\sigma_s(\mu\sin\theta + \cos\theta)}{2(1 - v^2)(\mu t/2 + h'/\sin\theta)} = \frac{wt^2\sigma_s(\mu\sin\theta + \cos\theta)}{2(1 - v^2)\left\{\mu t/2 + \left[h - \left(r_p + r_d + t\right)(1 - \cos\theta)\right]/\sin\theta\right\}} \qquad (17)$$

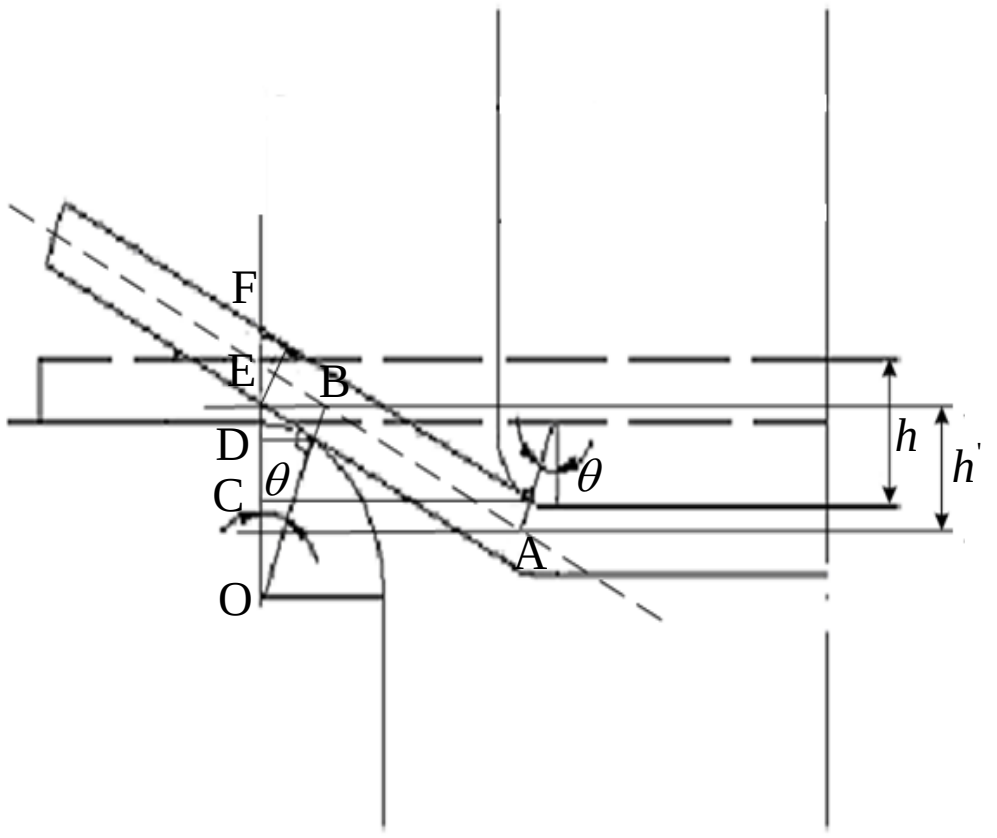

Fig.4 Geometrical relationship of $h < r_p + r_d + t$

(2)When $h \geq r_p + r_d + t$, according to the geometrical relationship, as shown in the Fig.1, the following can be obtained:

$$h' = h - \left(r_p + r_d + t\right)(1 - \cos\theta) = S_1 \sin\theta$$

$$h = S_1 \sin\theta + [r_d - (r_d + \frac{t}{2})\cos\theta] + [r_p - (r_p + \frac{t}{2})\cos\theta] + t \qquad (18)$$

The sheet enters into the elastic-plastic stage with the proceeding of bending, the value of M could be expressed by the equation (4), the bending moment gradually enlarge with the increase of bending stroke till the maximum value (that is the bending moment at point A). For the case of the larger ratio of $r_i/t$, the bending force can be expressed as:

$$p = \frac{\frac{KR^{n+1}}{2+n} wt^2 \left(\frac{t}{2r_p'}\right)^n \left(1 + \frac{3}{2}\frac{n+2}{n+3}\frac{t}{2r_p'}\right)(\mu\sin\theta + \cos\theta)}{\left(\mu t/2 + S_1\right)} \qquad (19)$$

## 3 Experimental verification

The device of bending test is shown in the Fig.5.

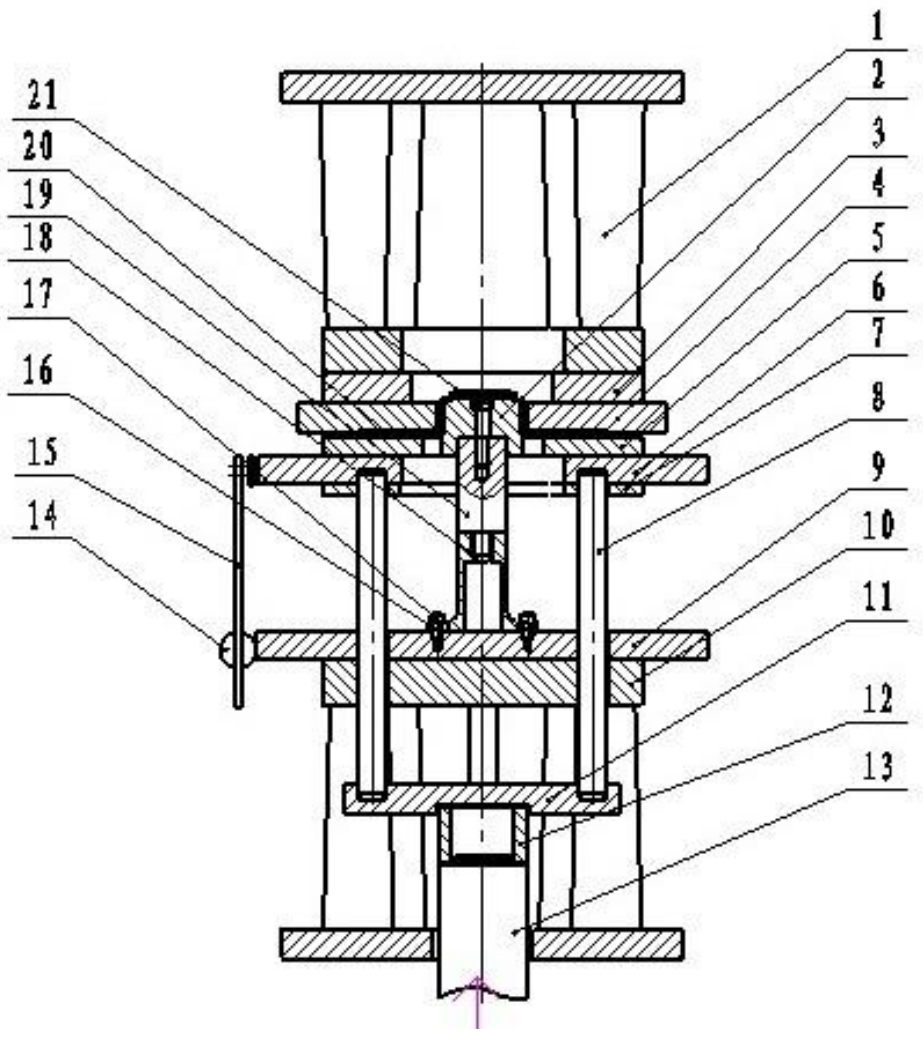

1 upper shoe,2 male punch,3 die holder,4 lower die,5 blank holder,6 blank holder fixed plate,7 ejector locating plate,8 ejector pin,9 bending force sensor fixed plate,10 die shoe,11 upper plate,12 pressure-pad-force senor,13 ejector piston extension rod,14 displacement sensor,15 displacement amplification sloping block,16 pressure ring,17 screw,18 bending force sensor,19 punch connector,20 bool,21 socket head screw

Fig.5 Schematic of bending apparatus

In order to use the hydraulic ejector cylinder to provides pressure-pad-force, a flip-bending die structure is adopted. The male punch 2 is connected by the punch connector 19 and the bending force sensor 18, and fixed together on the die shoe 10 by the pressure ring 16. The lower die 4 and the upper shoe 1 drop with the press movable beam, and force the blank holder 5 to decline simultaneously. The pressure-pad-force provided by the ejector cylinder transfer to the blank holder 5 through the ejector piston extension rod 13, the pressure-pad-force senor 12, the upper plate 11 and the ejector pin 8. The displacement sensor 14 is transversely placed on the side of the die shoe, the displacement amplification sloping block 15 converts vertical displacement of the press to horizontal displacement of the displacement sensor, at the same time enlarges the measuring range of the displacement sensor in proportion.

The shapes of the plate under different conditions after rebound in experiment are shown in the Fig.6. The results of the experiment and the numerical values from the theoretical molds are compared and analyzed.

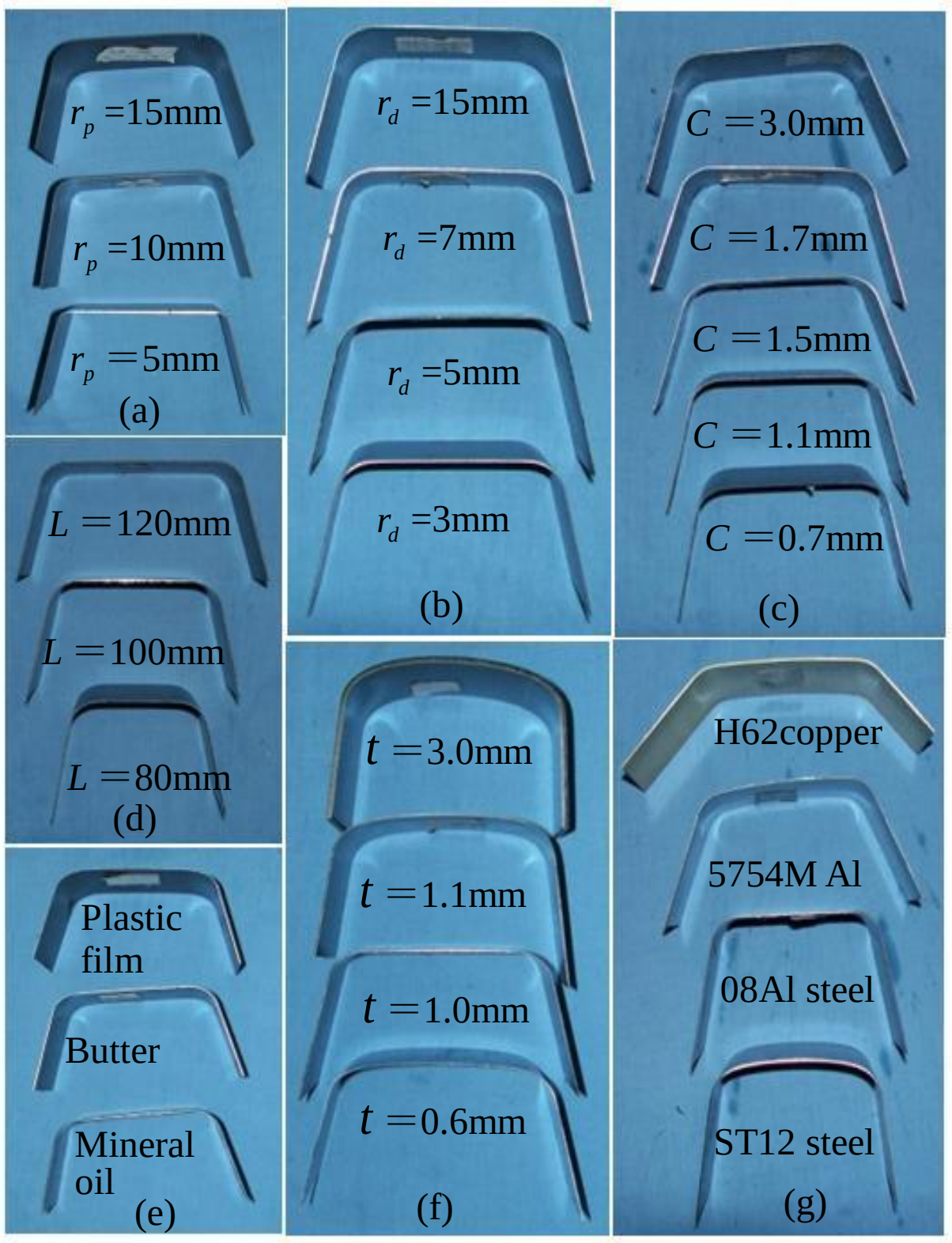

Experiment condition: (a)the material is 08Al steel plate, die span $L$=100mm, the clearance between punch and die $C=1.1t$, die profile radius $r_d$ =3mm, plate thickness $t$ =0.9mm, hydraulic oil lubrication, different punch profile radius $r_p$; (b)the material is 08Al, $L$=100mm, $C$ = 1.1 $t$ , $r_p$ = 15mm , $t$ = 0.9mm, hydraulic oil lubrication, different punch profile radius $r_d$ ; (c)the material is 08Al, $L$=100mm, $r_d$ =3mm, $r_p$ =15mm, $t$ =0.9mm, hydraulic oil lubrication, different clearance between punch and die $C$; (d)the material is 08Al, $r_p$ =15mm, $r_d$ = 3mm, $C$ =1.1$t$, $t$ =0.9mm, hydraulic oil lubrication, different die span L; (e)the material is 08Al, $C$ =1.1$t$, $r_p$ =15mm, $r_d$ =3mm, $t$ =0.9mm, $L$=100mm, different lubricant lubrication; (f)the material is 08Al, $C$ =1.1$t$, $r_p$ =15mm, $r_d$ =3mm, $L$=100mm, hydraulic oil lubrication, different plate thickness $t$; (g) $C$ =1.1$t$, $r_p$ =15mm, $r_d$ =3mm, $L$=100mm, $t$ =0.9mm, hydraulic oil lubrication, different plate material.

Fig.6 Shapes of the plate under different conditions after rebound in experiment

## 4 Outcome and Discussion

### 4.1 Factors affecting bending force

From the equations (17) and (19), one can see that the factors affecting bending force are material property (hardenability value $n$, yield strength $\sigma_s$, strength factor $k$, Anisotropy coefficient $R$), physical dimension of mold and sheet (punch profile radius $r_p$, die profile radius $r_d$, clearance $c$, material thickness $t$, plate width $w$), bending stroke $h$, friction coefficient $\mu$ [9].

The relationship between bending force $P$ and bending stroke $h$ of the sheet U-free bending is shown in the Fig.7(the material is 08Al, yield strength $\sigma_s$ = 250MPa、strength factor $k$ =511.81MPa、hardenability value $n$ =0.19、elasticity modulus $E$ =156.436GPa、Anisotropy coefficient $R$ =1.62; the clearance between punch and die $c$ =1.0mm, punch profile radius $r_p$ = 15mm, die profile radius $r_d$ =3mm, plate thickness $t$ = 0.9mm). It can be seen from the figure that there is a certain error between theory and experimental measurement, the maximum relative error is 10%, but the variation tendency is the same, it proved that using theoretical mold to analyze wide plate U-bending is feasible.

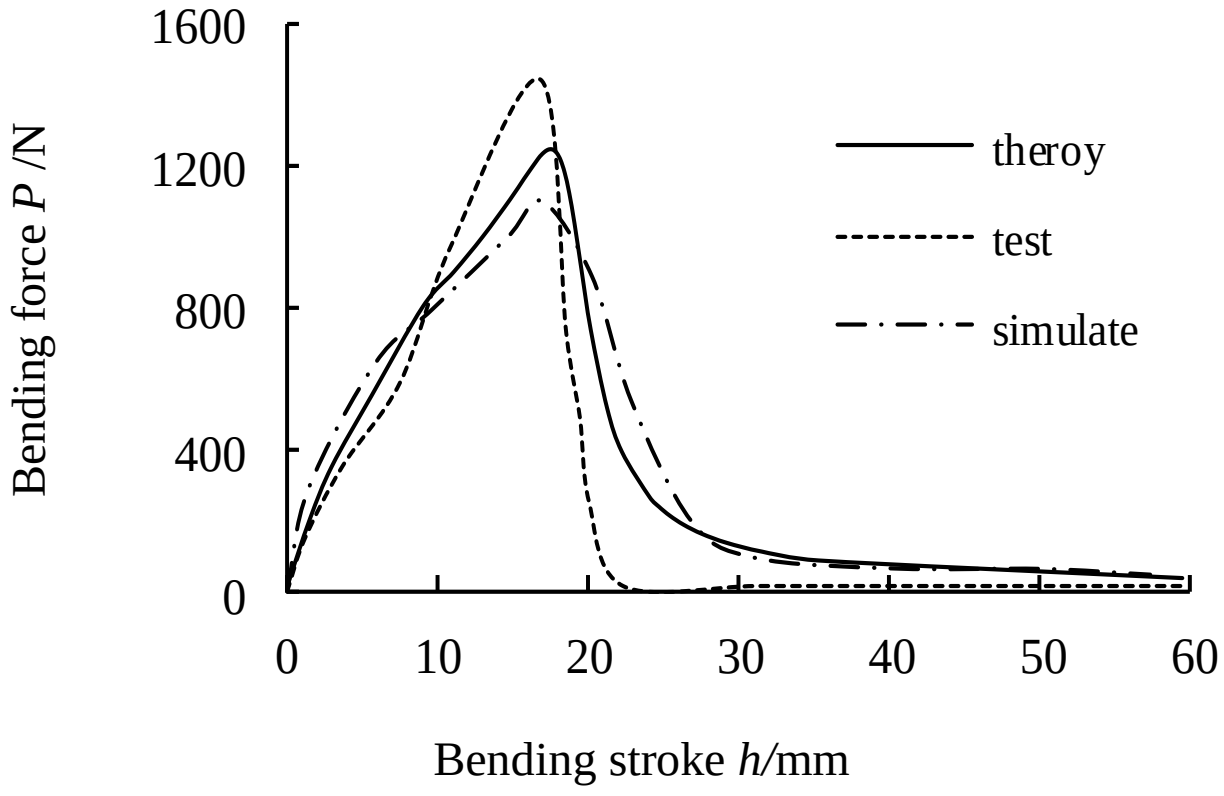

Fig.7 Relationship of bending stroke and bending force

1) **Influence of the clearance between punch and die on bending force** As shown in the Fig.8, influence on bending force is researched only by changing the clearance $c$ between punch and die and without changing other conditions. From the figure, with the increase of the clearance between punch and die, the maximum bending force decreases.

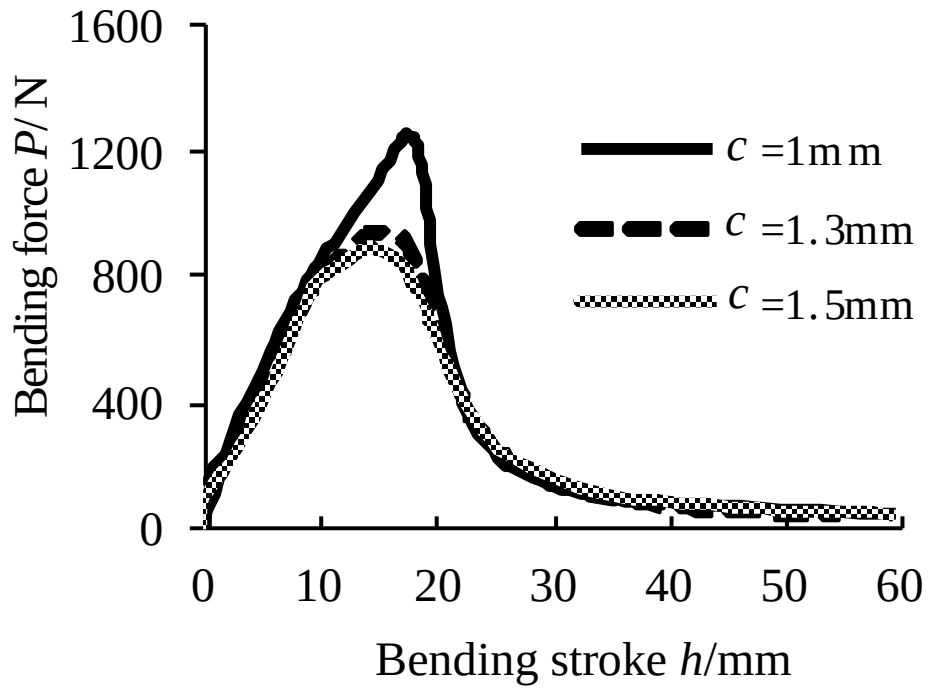

Fig.8 Influence of different clearances on bending force

2) **Influence of punch profile radius on bending force** Influence of different punch profile radiuses $r_p$ on bending force is shown in the Fig.9. From the figure, with the increase of punch profile radius, the maximum bending force decreases, and the trend is relatively obvious.

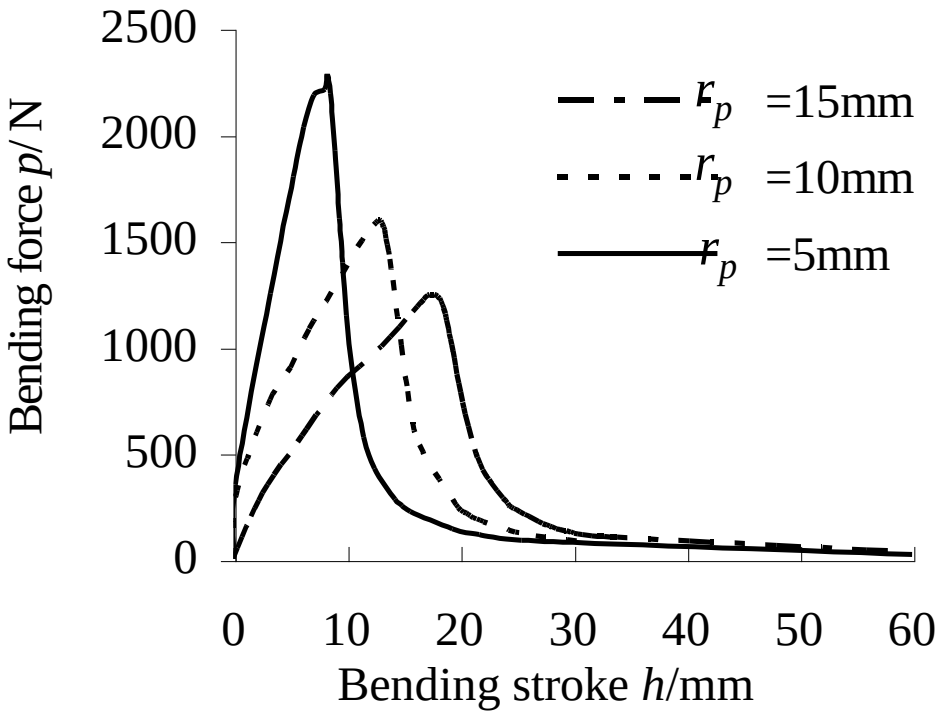

Fig.9 Influence of punch profile radius on bending force

3) **Influence of die profile radius on bending force** Influence of different die profile radiuses on bending force is shown in the Fig.10. From the figure, with the increase of die profile radiuses, the maximum bending force decreases.

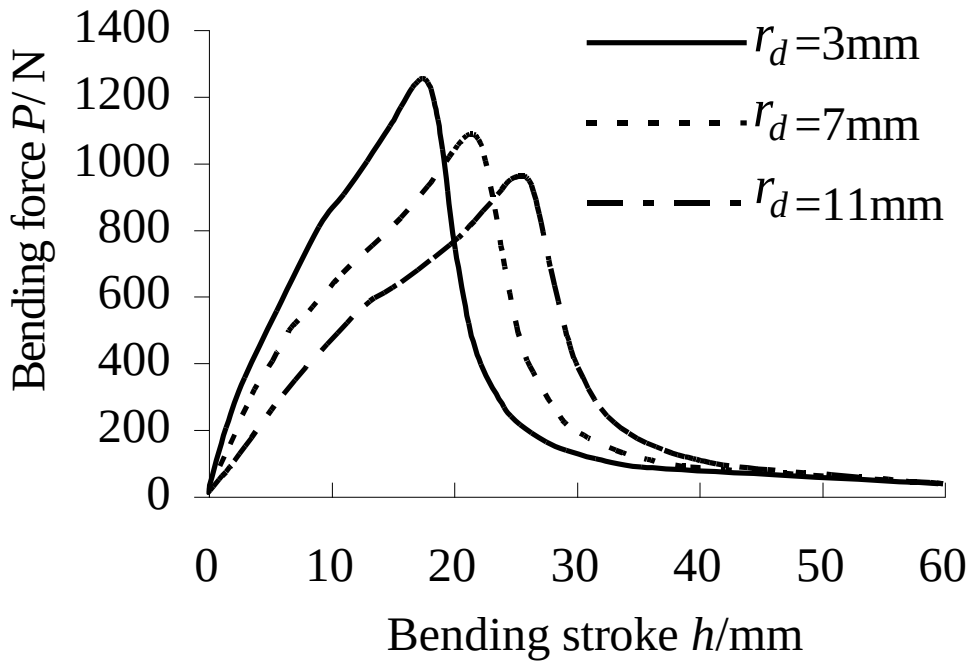

Fig.10 Influence of die profile radius on bending force

4) **Influence of plate thickness on bending force** Influence of different plate thicknesses $t$ on bending force is shown in the Fig.11. From the figure, with the increase of plate thickness, the maximum bending force decreases, the trend is very large, namely the influence of plate thickness on bending force is great.

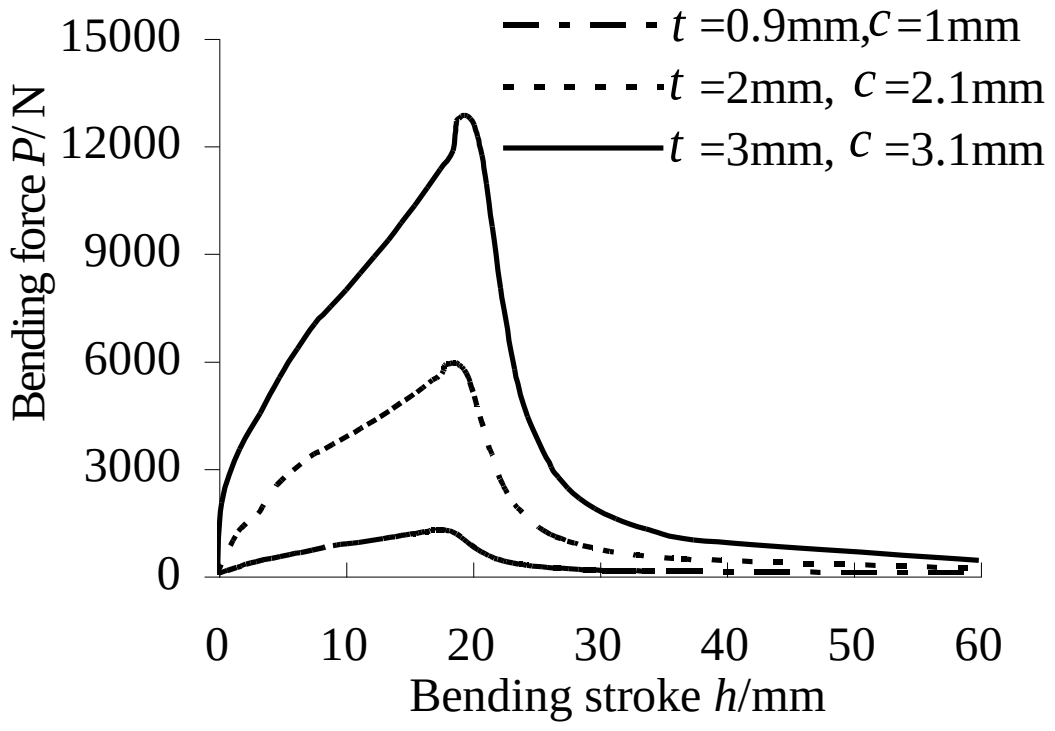

Fig.11 Influence of plate thickness on bending force

5) **Influence of friction coefficient on bending force** Influence of different friction coefficients on bending force is shown in the Fig.12. From the figure, with the increase of friction coefficient, the maximum bending force increases, but the influence is very small.

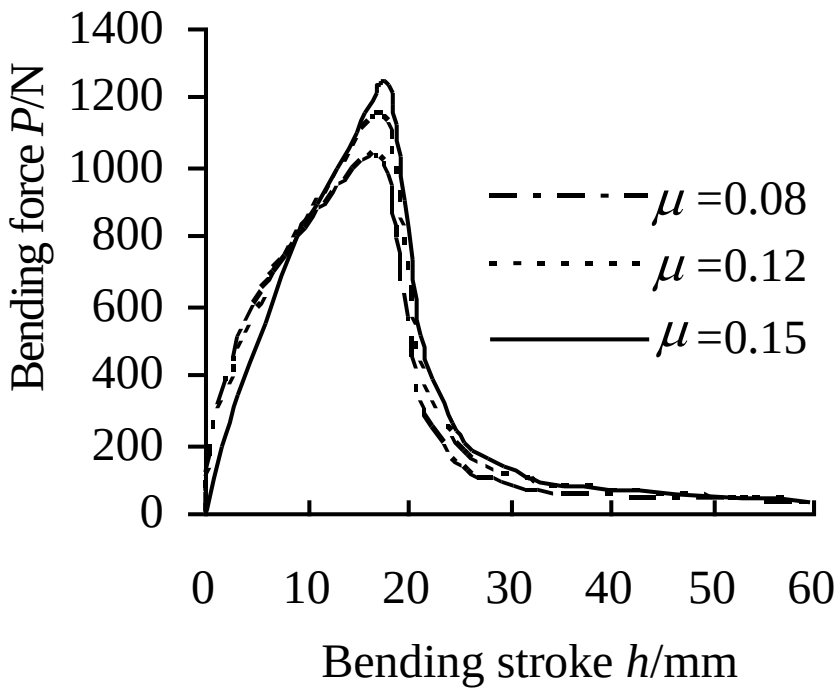

Fig.12 Influence of friction coefficient on bending force

6) **Influence of material property on bending force**

Influence of different material property on bending force is shown in the Fig.13, the material is 08Al steel plate, St12 steel plate and H62 copper plate(the yield strength of St12 is $\sigma_s$ =170MPa, strength factor $k$ = 493.62MPa, hardenability value $n$ =0.224 and Anisotropy coefficient $R$ =1.796; the yield strength of H62 is $\sigma_s$ = 445MPa, strength factor $k$ = 664.6MPa, hardenability value $n$ =0.109 and anisotropy coefficient $R$ =0.864). From the figure, the maximum bending force of H62 is the greatest, St12 followed, 08Al is the smallest.

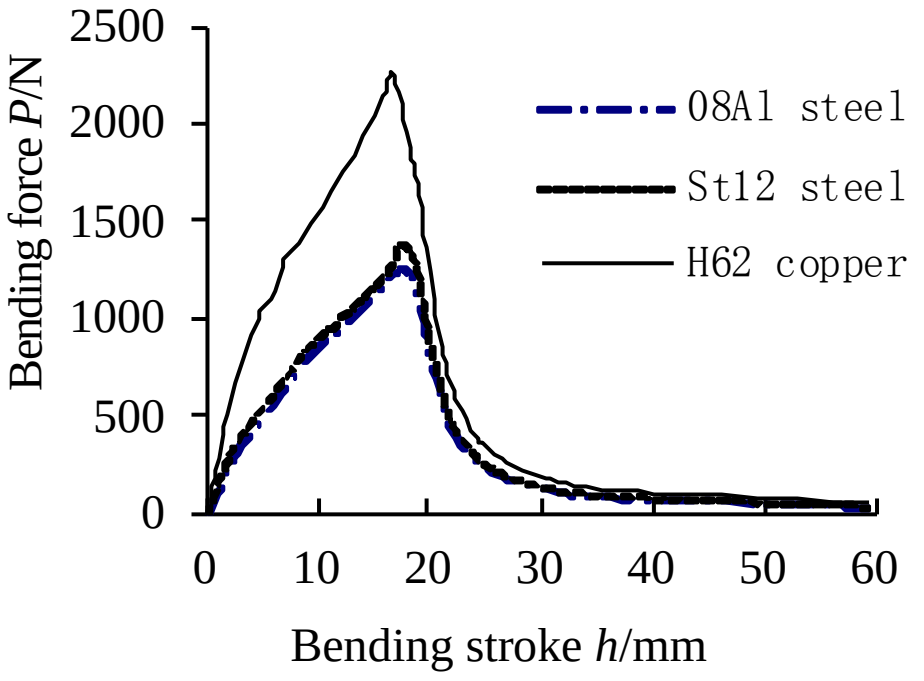

Fig.13 Influence of material on bending force

## 4.2 Factors affecting rebound

From the equation (12), we can see that factors affecting rebound angle include material performance parameter (strength factor $k$, hardenability value $n$, elasticity modulus $E$ and anisotropy coefficient $R$), mold dimension(punch profile radius $r_p$, die profile radius $r_d$ and the clearance $c$ between punch and die), bending stroke $h$ and plate thickness $t$.

Under the following conditions, analysis on the equation (12) is made: the material is 08Al, yield strength $\sigma_s$ =250MPa, strength factor $k$ =511.81MPa, hardenability value $n$ =0.19, elasticity modulus $E$ =156.436GPa, anisotropy coefficient $R$ =1.62, the clearance $c$ =1.0mm, punch profile radius $r_p$ =15mm, die profile radius $r_d$ = 3mm, plate thickness $t$ = 0.9mm

**1) Influence of plate thickness on rebound** Influence of plate thickness on rebound, while $\theta = 75°$, is shown in the Fig.14. From the figure, one can see that rebound declines with the increase of plate thickness, which is conformed with numerical simulation and experimental result. The maximum error between theory and experiment is 2 °, the maximum error between theory and numerical simulation is 5 °.

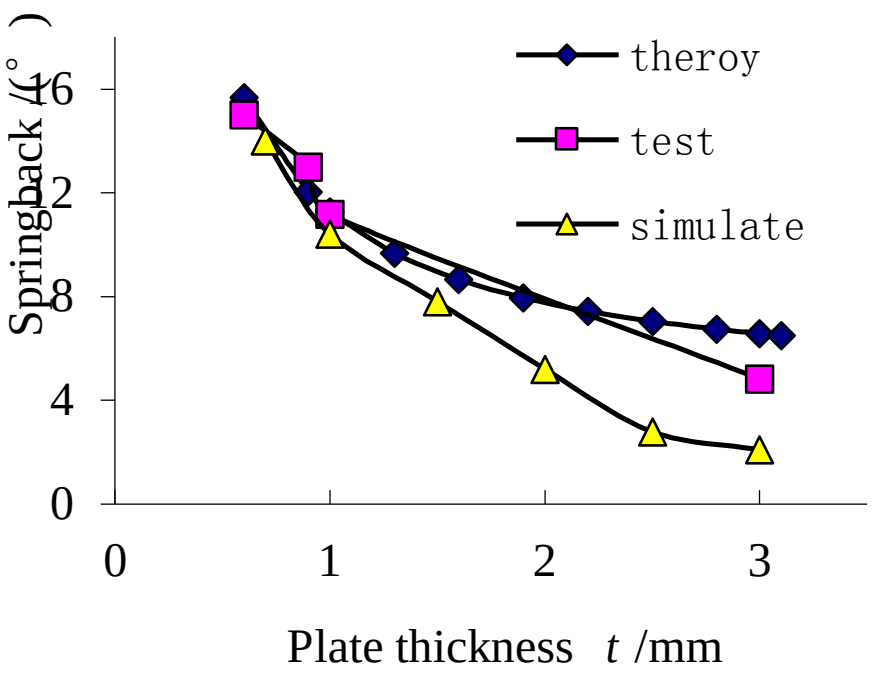

Fig.14 Influence of plate thickness on rebound angle

**2) Influence of punch profile radius on rebound angle** Influence of punch profile radius on rebound, while $\theta = 75°$, is shown in the Fig.15. From the figure, one can see that the rebound increases with the increase of punch profile radius, and the theoretical value is conformed with numerical simulation and experimental result. The maximum error is 2 ° within the range of $r_p \leq$ 20mm.

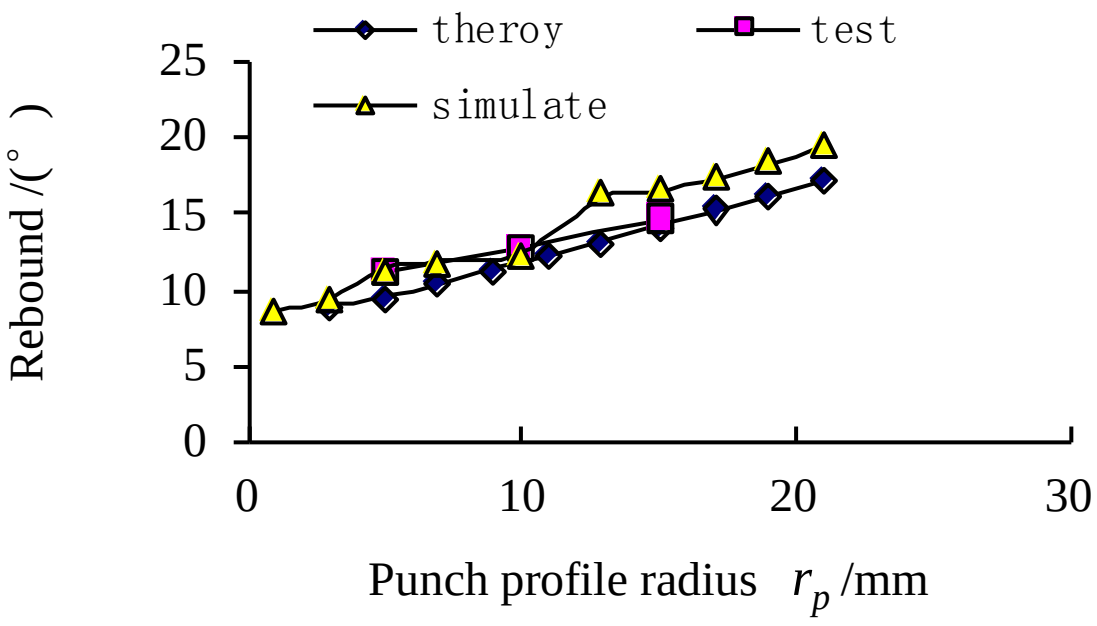

Fig.15 Influence of punch radius on rebound angle

**3)Influence of die profile radius on rebound** Influence of die profile radius on rebound, while $\theta = 75°$, is shown in the Fig.16. From the figure, one can see that the rebound increases with the increase of die profile radius, the numerical simulations and the experimental results are in accordance with the trend of the theoretical result. However, the error between theoretical calculation and experiment and the error between theoretical calculation and numerical simulation gradually increase with the increase of die profile radius, and the theoretical values are greater than the experimental and numerical simulation values. The main reason of which is that the theoretical calculation does not take friction into account, generally speaking, increasing the friction can reduce the rebound of sheet bending.

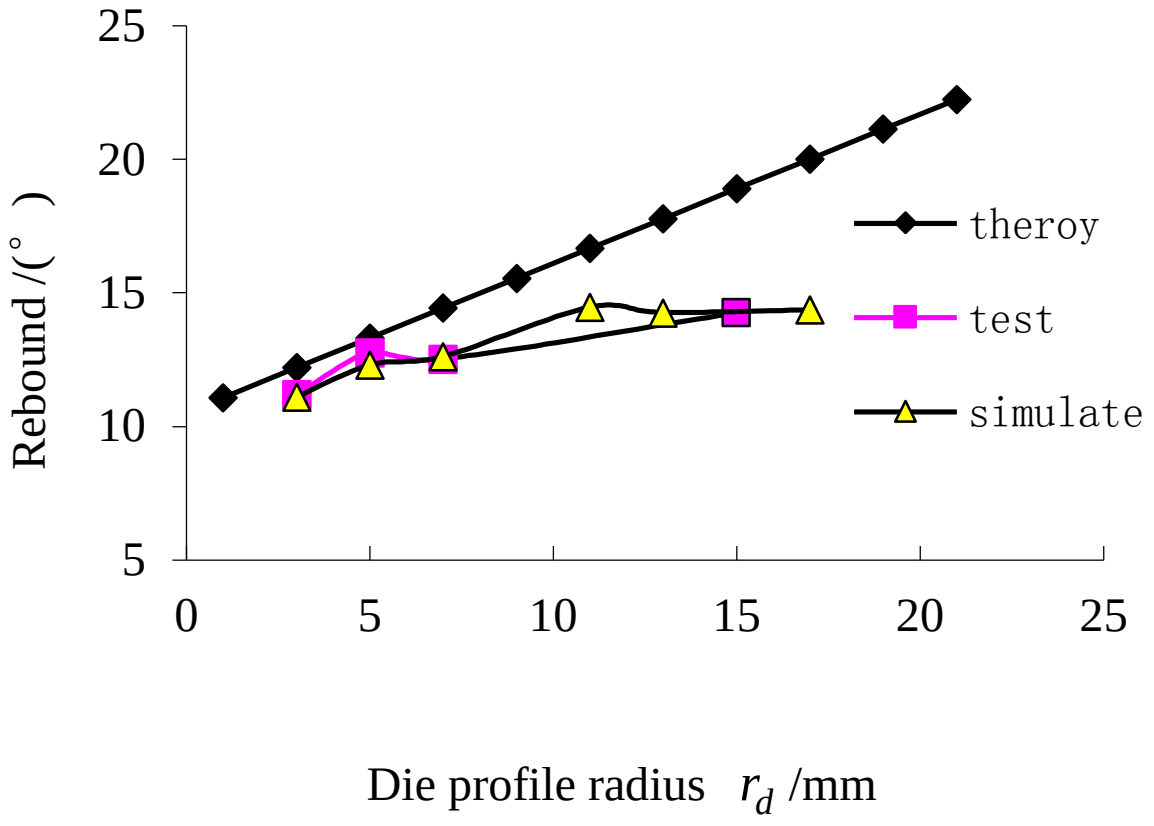

Fig.16 Influence of die radius on rebound angle

## 5 Conclusions

According to flat-bend theory, on the premise of plane deformation assumption, the harden, anisotropy and elastic deformation of material are considered, we analyzed the U-free bending of wide plate in theory, and obtained the change rule of bending force with bending stroke and formula of target bending angles. And we also analyzed the various factors affecting bending force and rebound. The major factors affecting bending force include punch profile radius $r_p$, plate thickness $t$ and material property. The factors which have little influence on affecting bending force include die profile radius $r_d$ and friction coefficient $\mu$, etc.

The rebound increases with the increase of strength factor $k$, anisotropy coefficient $R$, die profile radius $r_d$, punch profile radius $r_p$ and decreases with the increase of elasticity modulus $E$, Poisson ratio $v$ and plate thickness $t$. Among them the plate thickness $t$ and the clearance $c$ between punch and die have a greater influence on the rebound.

Although there is a large error in the theoretical calculation and numerical simulation and experiment, but the trend of the theoretical calculation and numerical simulation and experimental results is basically the same, which can be used as theoretical basis for identifying parameters of the wide plate U-free bending intelligent control as well as determining variables of the input layer and output layer of prediction model.

**Acknowledgments**

The authors would like to thank the anonymous reviewers for their helpful suggestions to this paper. This work was supported by the Postdoctoral Innovation Foundation of Shandong Province in China(201103095), Institutions of higher learning scientific research projects of Shandong Province in China(J12LA03) and Visiting scholar project communication plan of Shandong Province in China (Gf2013001).